\documentclass[aps,pra,groupedaddress,twocolumn,showpacs]{revtex4-1}
\usepackage{graphicx}
\usepackage{pstricks}
\usepackage{appendix}

\begin{document}
\title{Adiabatic Splitting, Transport, and Self-Trapping \\
of a Bose-Einstein Condensate in a Double-Well Potential}
\author{C.~Ottaviani,$^{1}$ V.~Ahufinger,$^{1,2}$ R.~Corbal\'{a}n,$^{1}$ and J.~Mompart.$^{1}$}

\affiliation{$^1$Departament de F\'{\i}sica, 
Universitat Aut\`{o}noma de Barcelona, E-08193 Bellaterra, Spain} 
\affiliation{$^2$Instituci\'{o} Catalana de Recerca i Estudis Avan\c{c}ats (ICREA), Llu\'{i}s Companys 23, E-08010 Barcelona, Spain} 

\begin{abstract}

We show that the adiabatic dynamics of a Bose-Einstein condensate (BEC) in a double well potential can be described in terms of a dark variable resulting from the combination of the population imbalance and the spatial atomic coherence between the two wells. By means of this dark variable, we extend, to the non-linear matter wave case, the recent proposal by Vitanov and Shore [Phys. Rev. A \textbf{73}, 053402 (2006)] on adiabatic passage techniques to coherently control the population of two internal levels of an atom/molecule. We investigate the conditions to adiabatically split or transport a BEC as well as to prepare an adiabatic self trapping state by the optimal delayed temporal variation of the tunneling rate via either the energy bias between the two wells or the BEC non-linearity. The emergence of non-linear eigenstates and unstable stationary solutions of the system as well as their role in the breaking down of the adiabatic dynamics is investigated in detail. 

\end{abstract}

\date{\today }
\maketitle

\section{Introduction}

Bose Einstein condensates (BEC) in double well potentials have drawn a lot of attention both theoretically and experimentally for the possibilities they offer to study fundamental quantum mechanical effects at the macroscopic level as well as for potential applications like interferometry, high precision measurements or thermometry  \cite{split}. Most of the experimental realizations lead to the splitting of the condensate into two independent parts. Nevertheless, recently, also weakly linked parts of a BEC in a double well potential forming a single Josephson junction \cite{josephson} have been achieved \cite{BEC_single_josephson}. In contrast to Josephson junctions realized in superconductors and superfluids \cite{classical_josephson}, in BEC the non-linear interatomic interactions play a crucial role. In the presence of the non-linearity two dynamical regimes have been predicted: (i) anharmonic Josephson oscillations \cite{anharmonic}, if the initial population imbalance of the two wells is below a critical value and (ii) macroscopic quantum self-trapping \cite{self-trapping} i.e, inhibition of large amplitude Josephson oscillations above the threshold for the population imbalance. This threshold corresponds to the population imbalance for which the difference between the two on-site interaction energies 
becomes larger than the tunneling energy splitting. Both dynamical regimes have been explored experimentally in a 
single Josephson junction \cite{BEC_single_josephson} and in arrays \cite{josephson_arrays}.

Recently, a lot of attention has been devoted to explore techniques to coherently control the non-linear dynamics of a BEC in double well potentials  by modulating in time either the potential \cite{time_potential} or the non-linearity \cite{time_nonlinearity}. 
The two main studied regimes are the non-linear Landau-Zener
\cite{WuNiu00,Liu03,The06,Gra06,Wan06,Iti07,Zha08} and the Rosen-Zener  
\cite{DYe08} regimes. In the former the tunneling rate is fixed while the energy bias is linearly varied in time, in the later the energy bias is varied in time while the tunneling rate is switching on/off following e.g., a temporal Gaussian profile. Both these regimes have been deeply investigated in double 
\cite{WuNiu00,Liu03,The06,Iti07,Zha08,DYe08} and triple well \cite{Gra06,Wan06} potentials yielding a wide variety of dynamical scenarios ranging from robust population transfer to quantum blocking, non-linear oscillations and, even, breaking down of the adiabatic dynamics.     

Following a different perspective, there have been several recent proposals to coherently manipulate single atoms \cite{NosPRA04,NosOC06,Das08} and BECs \cite{Gra06,Rab08,Nesterenko08} in triple-well potentials by adiabatically following a particular energy eigenstate of the system, the so-called spatial dark state. This spatial dark state only involves the two ground states of the extreme traps in a close analogy to the well known quantum optical Stimulated Raman Adiabatic Passage (STIRAP) technique \cite{Bergmann1998RMP70}. Accordingly, these tools were named Three-Level Atom Optics (TLAO) techniques \cite{NosPRA04}. 
In this paper, following the work by Vitanov and Shore \cite{Vitanov2006PRA73} on the adiabatic passage on two-level atoms, we extend the TLAO techniques to the two-level matter wave case showing that the adiabatic dynamics of a BEC, in a double well potential, can be described in terms of a dark variable resulting from the combination of the population imbalance and the spatial atomic coherence. This dark variable can be tailored varying in time the tunneling rate, the energy bias, and the non-linearity with the goal of (i) adiabatically splitting of the BEC, (ii) achieving complete BEC transfer from one well to the other; and (iii) preparing of an adiabatic self-trapping state. 

The paper is organized as follows. In Section~II we present the physical system consisting of a BEC in a double-well potential. We assume the two-level approximation and describe the BEC dynamics in terms of a dark variable. The conditions for the adiabatic control of tunneling by means of this dark variable are discussed in Section~III from a non-linear dynamics perspective. In Section~IV we present detailed numerical simulations on the adiabatic splitting, transport, and trapping of a BEC. Section~V summarizes the main conclusions of the paper and briefly discusses the validity of the two-mode approximation. It also presents some possible extensions of the present work such as its formulation in second quantization.

\section{Model}
We consider a BEC trapped in a double well potential (Fig.~1(a)), whose dynamics at zero temperature is described by the time-dependent Gross-Pitaevskii equation (GPE) ($\hbar =1$):
\begin{equation}
i \frac{d\psi (\vec{r},t)}{dt}=\left[ -\frac{\triangle}{2m} +V(\vec{r},t)+g|\psi (\vec{r},t) |^{2}\right] \psi (\vec{r},t) ,  \label{GPE}
\end{equation}%
where $V(\vec{r},t)$ is the external trapping potential and the non-linearity is given by $g=4N\pi a_{s}/m$, with $N$ the total atom number, 
$a_{s}$ the $s$-wave scattering length, and $m$ the atomic mass. 
Assuming the two-level approximation \cite{self-trapping,two_lev_approx}, the wave function or classical order parameter of the BEC under study can be written as 
$\psi (\vec{r},t)= c_L(t) \phi_L(\vec{r})+c_R (t)\phi_R(\vec{r})$ with $c_{L(R)}(t)=\sqrt{N_{L(R)}}e^{i\theta_{L(R)}(t)}$ satisfying  $\left|  c_L \right|^2 + \left|  c_R \right|^2=1$. $\phi_{L,R}(\vec{r})$ accounts for the ground state of the corresponding isolated trap. The amplitudes $c_L$ and $c_R$ obey the non-linear two-mode dynamical equations given by:
\begin{equation}
\mathcal{H} 
\left( 
\begin{array}{c}
c_L \\ 
c_R
\end{array}
\right) = i  {d \over {dt}}
\left( 
\begin{array}{c}
c_L \\ 
c_R
\end{array}
\right)
\end{equation}
with the Hamiltonian:
\begin{equation}
\mathcal{H}
= 
\left( 
\begin{array}{cc}
  \epsilon_{L}+ U_L \left|  c_L \right|^2  & \Omega \\ 
\Omega &  \epsilon_{R} + U_R \left|  c_R \right|^2   
\end{array}
\right)
\end{equation}
where $U_{R,L}$ are the atomic self-interaction energies, $\Omega$ is the tunneling rate between the two wells, and $\epsilon_{R,L}$ are the on-site energies. In terms of the $\phi_{L,R}(\vec{r})$ overlaps, the former parameters read:
\begin{eqnarray}
\Omega&=&-\int d\vec{r}\left[\frac{1}{2m}\nabla \phi_L^* \nabla \phi_R + \phi_L^* V(\vec{r},t)\phi_R \right]\\
U_{L,R}&=&g \int d\vec{r} \left|\phi_{L,R}\right|^4\\
\epsilon_{R,L}&=&\int d\vec{r}\left[\frac{1}{2m}\left|\nabla \phi_{R,L}\right|^2 +\phi_{R,L}^*V(\vec{r},t)\phi_{R,L} \right]
\end{eqnarray}
with $\int d\vec{r}[\phi_i^* \phi_j]=\delta_{ij} \, (i,j=L,R)$.

Notice that the two-mode approximation assumes that the parameters of the problem (tunneling rate, non-linear interaction, and energy bias) can be varied independently. In general, however, the temporal modification of any of these parameters will result in the modification of the spatial mode functions affecting the rest of the parameters. From the experimental point of view, a double-well optical potential can be created with well separations on the range of few microns by means of two focused laser beams \cite{dw1}, or by the superposition of a 3D crossed beam dipole trap with a 1D optical lattice \cite{BEC_single_josephson}. In the later, the intensity of the standing wave field that creates the optical lattice and its displacement with respect to the center of the dipole trap could be used to manipulate both the tunneling and the energy bias. In addition, the scattering length could be controlled playing around of a Feshbach resonance \cite{Bauer09}.

Within the density matrix formalism, assuming $U_R \sim U_L (\equiv U )$, and taking $\epsilon \equiv \epsilon_R-\epsilon_L$ as the energy bias between the two wells, the coherent dynamics of the BEC in the two-well potential can be written as follows \cite{LeeHai}:
\begin{equation}
\frac{d}{dt}\left( 
\begin{array}{c}
u \\ 
v \\ 
w
\end{array}
\right) =\left( 
\begin{array}{ccc}
0 & -\left( \epsilon + U w \right) & 0 \\ 
\left( \epsilon + U w \right) & 0 & -2\Omega \\ 
0 & 2\Omega & 0
\end{array}
\right) \left( 
\begin{array}{c}
u \\ 
v \\ 
w
\end{array}
\right) \label{Eqs_2L}
\end{equation}
where we have introduced the real-valued variables  
$
u=2
\mathop{\rm Re}
\left\{ \sigma _{LR}\right\}$, $v=2
\mathop{\rm Im}
\left\{ \sigma _{LR}\right\}$, ~$ w= \sigma _{RR}-\sigma
_{LL} $
being $\sigma _{ii}=c_ic^*_i$ and $\sigma_{ij}=c_ic^*_j$ with $i,j=L,R$ the corresponding trap populations and spatial coherence (see Fig.~1(b)).
Note that the conservation of the norm implies $w^2+u^2+v^2=1$. 

\begin{figure}[t]
\includegraphics[scale=1.5]{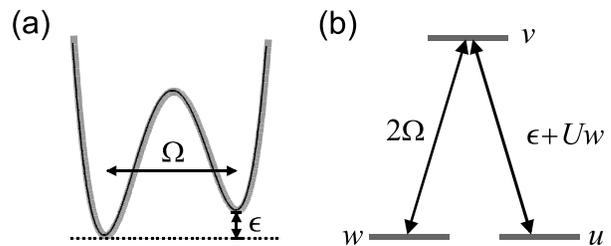}
\begin{picture}(0,0)(0,0)
\end{picture}
\caption{\label{fig:scheme} (a) Two-level model for the BEC in a double-well potential being $\Omega $ the tunneling rate and $\epsilon $ the energy bias. 
(b)~Three-level correspondence of (a) for the density matrix variables and coupling strengths. $w$ is the population difference, $u/2$ ($v/2$) is the real (imaginary) part of the spatial coherence, and $U$ is the non-linear self-interaction energy. ($\hbar = 1$)
}
\end{figure}

Inspired by the work of Vitanov and Shore \cite{Vitanov2006PRA73} on STIRAP in a system of two internal atomic levels, we extend their proposal to the external degrees of freedom of matter waves. It is easy to verify that Eqs.~(\ref{Eqs_2L}) are equivalent to the discrete non linear Schr\"{o}dinger equation (DNLSE) for a BEC in a triple well potential (see, for instance, Ref.~\cite{Gra06}) under the following identifications: $u\Leftrightarrow 
{C}_{R}$, $v\Leftrightarrow - i {C}_{M}$, $w\Leftrightarrow
{C}_{L}$, $\epsilon + U \omega \Leftrightarrow \Omega
_{MR} $ and $2\Omega \Leftrightarrow \Omega _{LM}$,
where $C_{i}$ with $i=L,M,R$
are the probability amplitudes to find the BEC in the left, middle, or right trap, respectively. 
$\Omega _{LM}$ and $\Omega_{MR}$ denote, respectively, the tunneling interaction between the left-middle and middle-right traps. 
Note however that for the two level system the variables $u,v$ and $w$ are real quantities,
while for a three level system the variables $C_{i}$ are, in general, complex numbers.

In the TLAO techniques within the three-level approximation being the left and right levels resonant, the transfer process is based on adiabatically following one of the three energy eigenstates of the system 
 $\left| D  (\Theta) \right> = \cos \Theta  \left| L \right> - %
\sin \Theta  \left| R \right> $, with the mixing angle defined as
$\Theta ( t ) = \tan^{-1} \left[ \Omega_{LM}(t) / \Omega_{MR}(t) \right]$, and 
$\left| L \right>  $ and $\left| R \right> $ being the ground states of the left and right wells.
In the two-level system under investigation, the analogue state will be given by  
a combination of the population difference and the coherence
$ d (\theta) = \cos \theta  \times w - \sin \theta  \times u$,
with the mixing angle given by 
$\theta ( t ) = \tan^{-1} \left[ 2\Omega (t)  / (\epsilon (t) + U (t) w (t) ) \right]$.
This analogy opens the possibility to extend the TLAO techniques
to two-level systems, by appropriately engineering the time dependence of the tunneling rate $\Omega (t)$, the energy bias $\epsilon (t)$, and/or the non-linear interaction $U (t)$.  
Note that the non-linear interaction parameter can be modified in time by the temporal variation of the scattering length $a_s$ using either magnetic \cite{magnetic_FR} or optical \cite{optical_FR} Feshbach resonances or varying the trap frequency, leading to a modification of the BEC spatial profile according to Eq.~(5). For a review on the manipulation of Feshbach resonances see \cite{ReviewFeshbach}.

\section{Adiabatic Control of Tunneling}

In this section we study different scenarios for the coherent control of the external degrees of freedom of a BEC by adiabatically following the dark variable $d(\theta)$. In particular, we show how to adiabatically split, transport, and inhibit tunneling of a BEC via the matter wave analogues of both STIRAP and double-STIRAP techniques in two-level systems.  

Let us assume that the BEC is initially prepared in the left trap with $\Omega (t=-\infty)=0$. If so, note then that $\theta = 0 $ meaning $w=-1$ and $u=v=0$ will be the initial state. Then, to coherently split the condensate, $\theta$ should vary adiabatically from $\theta = 0 $ to $\theta = \pi/2 $ corresponding to $u=1$ and $w=v=0$ (see Fig.~1(b)) by appropriately changing $\Omega$, $\epsilon$, and $U$ in time. To transfer the BEC from the left to the right trap, the mixing angle should be slowly increased up to $\theta = \pi$. For the adiabatic self-trapping state case, the evolution will consist in varying $\theta$ from $0$ to e.g., $\pi/2$ and then back to $0$. In all cases, in order to guarantee that the BEC follows the dark variable $d(\theta)$ during the whole process two conditions must be fulfilled:

\bigskip   
\noindent   
\emph{Condition 1: Adiabaticity criteria.} $\Omega (t)$, $\epsilon (t)$, and $U (t)$ should be smoothly varied in time to adiabatically follow the dark variable $d(\theta )$ which, in turn, means that the mixing angle $\theta(t) = \tan^{-1} 2 \left[ \Omega (t) /( \epsilon (t)  + U (t)  w ) \right]$ must be slowly changed from $\theta(t=-\infty)=0$ to its expected final value. Gaussian profiles for the temporal variations of the control parameters are assumed:
\begin{eqnarray}
\Omega &=& \Omega_{g}+\Omega_0
e^{{-(t-t_{\Omega})}^2 / {{ \sigma^2_{\Omega}}}} 
\\
\epsilon &=& \epsilon_{g}+ \epsilon_0
\left[
{ e^{-(t-t_{\Omega}+\Delta t_{{\epsilon}})^2 /  { \sigma^2_{{\epsilon}}}}+ } \right.  \nonumber \\
 & & \qquad \qquad \qquad \qquad \,  \left. {n_{\epsilon}  e^{-(t-t_{\Omega}- \Delta t_{\epsilon})^2 / { \sigma^2_{{\epsilon}}}}}
\right]
\\
U &=& U_{g}+ U_0
\left[ {e^{-(t-t_{\Omega}+ \Delta t_{U})^2 /  { \sigma^2_{U}}}+  } \right. \nonumber \\
 & & \qquad \qquad \qquad \qquad \,  \left. {n_U  e^{-(t- t_{\Omega}-\Delta t_U)^2 / { \sigma^2_{U}}}}
\right]
\end{eqnarray}
with either the energy bias $\epsilon (t)$, or the BEC non-linearity $U (t)$, preceding the tunneling interaction $\Omega (t)$, i.e., the counter intuitive sequences $\Delta t_{\epsilon}> 0 $ or $\Delta t_{U} > 0 $ will be assumed. 
$n_{{\epsilon},U}=0,\pm1$ is a switch that takes the value $0$ for the two-level matter wave analogue of the STIRAP sequence and $+1$ ($-1$) for the symmetric (antisymmetric) double-STIRAP sequences.
Adiabaticity means that, at any time, $\dot{\left| \theta (t) \right|}$ should be much smaller than the energy separation between the selected eigenstate and the one energetically closest. For a weak non-linear interaction, $\left| U (t) / \Omega (t) \right| \ll 1 $, this adiabaticity condition reads $ \dot{\left| \theta (t) \right|} \ll \sqrt{4\Omega (t)^2 + (\epsilon (t) + U (t) \omega )^2} $. 
It is worth to remark that for $U \neq 0 $, $\theta (t)$ is not a parameter of the system but a dynamical variable since it contains the population difference $w$ in its definition. 

\vskip 2cm 
\noindent 
\textit{Condition 2: Avoiding adverse bifurcation points.} For large enough values of the non-linearity, the interaction between the atoms of the BEC results in a non-linear temporal coupling producing additional non-linear stationary states yielding loop structures and a rich variety of level crossing scenarios \cite{Gra06,WuNiu00,TL2,TL3}. In the matter wave STIRAP case for a triple-well potential \cite{Gra06}, it has been shown that even in the adiabatic limit given by $ \dot{\left| \theta (t) \right|} \rightarrow 0$, the appearance of non-linear stationary states breaks down, in some cases, the adiabatic evolution.
To address this issue in the double-well potential, we start first looking for those critical $U$ values giving rise to non-linear stationary states by solving the eigenvalue equation: 
\begin{equation}
\mathcal{H} 
\left(
\begin{array}{c}
c_L \\ 
c_R
\end{array}
\right)
=\mu
\left(
\begin{array}{c}
c_L \\ 
c_R
\end{array}
\right)
\end{equation} 
with $\mathcal{H}$ given in Eq.~(3) and $\mu$ being the chemical potential. After some algebra one obtains the following fourth-order eigenvalue equation for $\mu$: 
\begin{eqnarray}
\left[ \epsilon^2 - \left( -2 \mu + \epsilon + 2 U \right)^2 
\right] \left( -2 \mu + \epsilon + U \right)^2 \nonumber \\
+ 4 \Omega^2 \left( -2 \mu + \epsilon + 2 U \right)^2 = 0
\end{eqnarray}
For $\epsilon = 0$ and $U^2/(2\Omega)^2 < 1$ ($> 1$) this quartic equation gives two (four) real roots \cite{WuNiu00}. 

Additional information on the BEC dynamics can be obtained looking for the stationary solutions of the density matrix equations (7) and analyzing their stability (see Appendix~A). The stationary solutions can be written as $\left\{ u^{ss},v^{ss}=0,w^{ss} \right\} $ with
$u^{ss}/w^{ss}=\tan \theta^{ss}$ and $(u^{ss})^2+(w^{ss})^2=1$. The energy of these stationary solutions, up to four, is given by Eq.~(12). Note that, as the dynamics is conservative, the eigenvalues of the linear stability matrix must satisfy $\Sigma_{i=1}^3 \lambda_i = 0 $ for each stationary solution. In fact, a Linear Stability Analysis (LSA)  around any of these solutions yields one eigenvalue equal to zero together with either two pure imaginary eigenvalues corresponding to a stable fixed point (or center) or two real eigenvalues accounting for an unstable saddle point. In the limit of a large non-linear interaction, i.e., for $U^2/(2\Omega)^2 > 1$, the system presents four stationary solutions, one of them being an unstable saddle point while the other three are stable elliptical ones. 

If the number of stationary solutions remains constant during the whole adiabatic dynamics, the system will successfully follow the selected energy eigenstate. However, depending on the interplay between the non-linear interaction and the tunneling rate, during the dynamics the number of stationary solutions will change through a bifurcation point from four to two or vice versa. If the selected stationary solution is not involved in the bifurcation, the system will again adiabatically follow the corresponding eigenstate. On the opposite case, whether the adiabatic dynamics will be affected will depend on the particular bifurcation scenario: 

\vskip 2cm 
\noindent \textsl{Case 1}. From 4 to 2 stationary solutions through a backward
pitchfork bifurcation point in which the stationary solution to be followed merges two more solutions (one unstable) and yields one stable solution. In this case, the system will evolve adiabatically. In what follows, this case will be termed PB42 case, even if the pitchfork bifurcation is perturbed and becomes imperfect. 

\smallskip
\noindent \textsl{Case 2}. From 2 to 4 solutions through a forward
pitchfork bifurcation point in which the stationary solution to be followed becomes unstable and yields two 
stable solutions corresponding, in the limit of vanishing tunneling, to $w = \pm 1$. In this case, the system after reaching the bifurcation point will split into a 
combination of the two stationary solutions. In this case, named PB24 in what follows, the adiabatic dynamics will, in general, break down.
  
\smallskip  
\noindent \textsl{Case 3}. From 4 to 2 solutions through a saddle-node bifurcation in which the stationary solution to be followed is annihilated by an unstable solution. The adiabatic dynamics breaks down even in the adiabatic limit. After reaching the bifurcation the system will split into a combination of two non-degenerate energy eigenstates and, therefore, will oscillate at the corresponding energy splitting. Case SNB42 in what follows.          
\section{Numerical Simulations}

This section is devoted to the numerical simulations of the BEC adiabatic splitting, transport, and self-trapping by means of the temporal variation of the energy bias, the non-linearity, and the tunneling rate. We will show, for the most relevant cases, the eigenvalues and stationary states predicted by Eqs.~(12) and (7) as well as their linear stability. The main goal of these numerical simulations will consist in illustrating the different dynamical scenarios to adiabatically control the BEC while looking for those parameter values that prevent the system to reach the two unwanted bifurcation cases PB24 and SNB42 previously described.

\begin{figure}[t]
\includegraphics[scale=1.2]{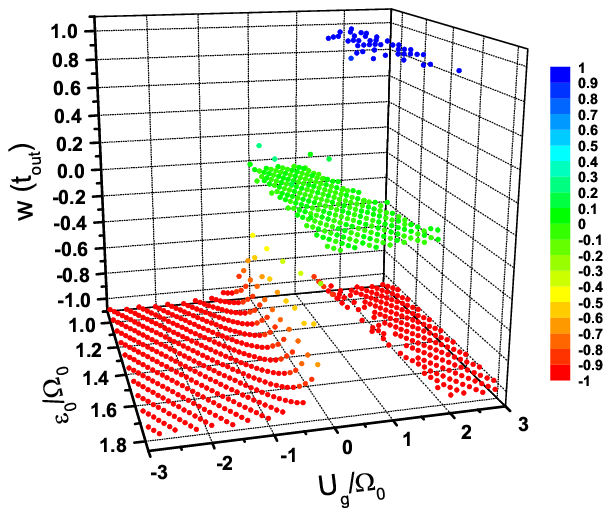}
\vskip 0.2cm
\includegraphics[scale=1.1]{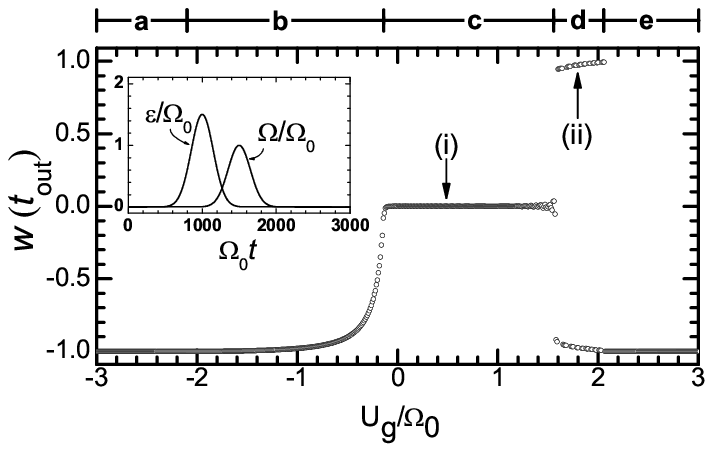}
\begin{picture}(0,0)(0,0)
\put(-230,310){{\bf (a)}}
\put(-230,132){{\bf (b)}}
\end{picture}
\caption{\label{fig:scheme} (Color online) Adiabatic splitting of a BEC. (a) Population difference $w(t_{out})$ at the end of the splitting process in the parameter plane  modulation amplitude of the energy bias $\epsilon_0$ versus the non-linear interaction parameter $U_g$. Initial conditions: $w(t_{in})=-1$ and $u(t_{in})=v(t_{in})=0$. Parameter setting: $ \sigma_{\epsilon}=\sigma_{\Omega}=212.8\Omega_0^{-1}$, $\Delta t_{\epsilon}=500\Omega_0^{-1}$, $t_{\Omega}=1500\Omega_0^{-1} $ and $\Omega_g=\epsilon_g=n_{\epsilon}=U_0=n_U=0$. 
(b) As in (a) for the fixed value $\epsilon_0=1.5\Omega_0$. 
The temporal variation of the energy bias and the tunneling rate is shown in the inset. From a non-linear dynamics perspective, we have classified the results into five different regions, from $a$ to $e$, depending on the number and type of bifurcations that the system suffers during its dynamics (see text). The detailed temporal dynamics corresponding to cases (i) and (ii) are plotted in Figs.~3 and 4, respectively. The adiabatic splitting succeeds in region $c$ corresponding to $U_g/\Omega_0 = (-0.12,1.5)$. 
}
\end{figure}

\begin{figure}[t]
\includegraphics[scale=1.2]{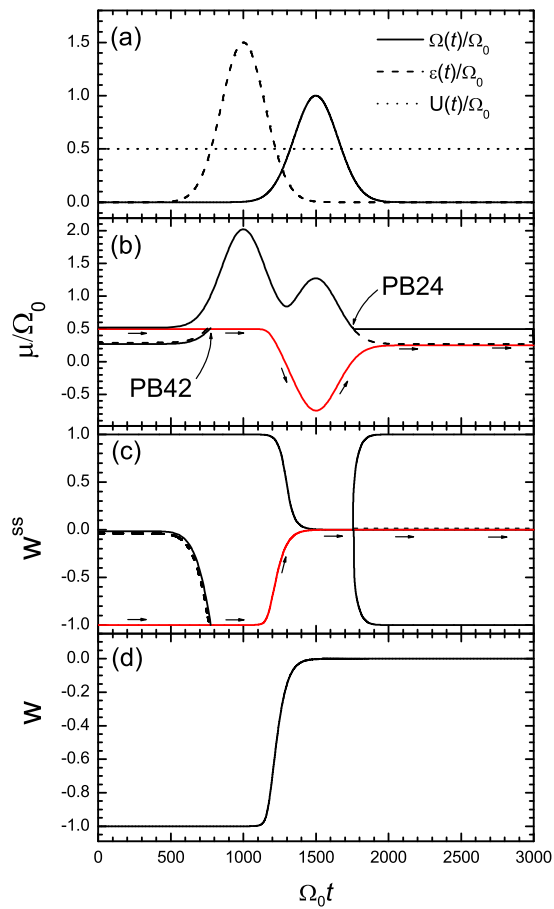}
\caption{\label{fig:scheme} (Color online) Adiabatic splitting of a BEC. $U_g=0.5\Omega_0$ and the rest of parameters as in Fig.~2. As a function of time: (a) non-linear interaction, energy bias and tunneling rate profiles; (b) energy eigenvalues; (c) stationary solutions for the population difference $w^{ss}$; and (d) population difference $w(t)$ after numerical integration of Eqs.~(7). In (b) and (c), the dashed curve accounts for the unstable solution, while the short arrows indicate the adiabatic solution (in red) to be followed by the system. PB42 and PB24 denote the two pitchfork bifurcations that the system suffers at $\Omega_0 t = 777$ and $\Omega_0 t = 1750$, respectively.
Note that the pitchfork bifurcation PB24 yields one unstable solution (dashed curve) plus two energetically degenerated stable solutions (solid curve) corresponding, in the limit of vanishing tunneling, to $w = \pm 1$.
 }
\end{figure} 

\subsection{Adiabatic Splitting of a BEC}

We start first by fixing the non-linear interaction control parameter $U$ to a constant value $U_g$ while, counter intuitively, time-varying the energy bias and the tunneling rate (see the inset of Fig.~2(b))) with the goal of achieving equal population in the two wells, i.e., $w (t_{out})=0$ starting with the BEC being in the left trap, i.e., $w (t_{in})=-1$, or, equivalently, $\theta (t_{out})=\pi /2$ from $\theta (t_{in})=0$. 
Fig.~2(a) shows $w (t_{out})$, after the numerical integration of Eqs.~(7) with $w(t_{in})=-1$ and the rest of parameters values given in the figure caption. Clearly, there is a large set of parameters where the adiabatic splitting process takes place with a high fidelity (see the plateau in Fig.~2(a)). 
Fig.~2(b) shows the plateau $w(t_{out})=0$ for $U_g/\Omega_0 = (-0.12,1.5)$ and $\epsilon_0 =1.5\Omega_0$.
To give more insight into the dynamics of the system, we have divided 
the non-linear interaction range studied in five regions, from $a$ to $e$, as shown at the top of Fig.~2(b). 
In the following we will discuss in detail the dynamics for each of these regions. 
Note that in all cases, the dynamics starts and ends with $\Omega (t_{in})=\Omega (t_{out}) =0$ which, for $U \neq 0$, implies that for $t=t_{in,out}$ the system presents four fixed points or stationary solutions. For intermediate times and large enough tunneling rates, there are only two stationary solutions. 

\begin{figure}[t]
\includegraphics[scale=1.2]{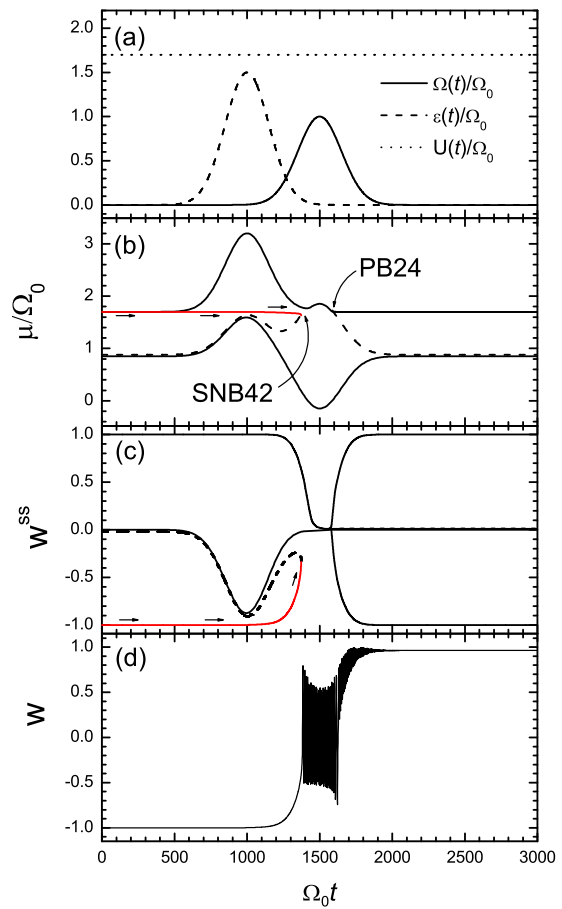}
\caption{\label{fig:scheme} (Color online) As in Fig.~2 for $U_g=1.7\Omega_0$. SNB42 and PB24 account for a saddle-node and a pitchfork bifurcation at $\Omega_0 t = 1374$ and $\Omega_0 t = 1588$, respectively.  
 }
\end{figure}

In Fig.~3 we show the temporal dynamics corresponding to the parameter values of case (i) in region (c) of Fig.~2(b) with $U_g=0.5\Omega_0 $ (see Fig.~3(a)). The adiabatic energy eigenvalues and stationary solutions $w^{ss}$ are shown in Figs.~3(b) and (c), respectively, with the dashed curve accounting for the unstable ones. Short arrows in Figs.~3(b) and 3(c) indicate the stationary solution to be adiabatically followed. The dynamical variable  $w(t)$, after integration of Eqs.~(7), is plotted in Fig.~3(d). For this set of parameter values the adiabaticity condition is fulfilled and the selected eigenstate is involved in a single bifurcation, at $\Omega_0 t=777$, corresponding to the PB42 case previously described. After the bifurcation point, the system follows the eigenstate that, at the end of the process, yields $w=0$ and $\mu=0.25\Omega_0$. This non-linear dynamical scenario holds for the whole plateau, region $c$ in Fig.~2(b), where the adiabatic splitting succeeds. 

For the parameter range $U_g/\Omega_0 = (1.5,2.07)$ corresponding to region $d$ in Fig.~2(b), the final state of the BEC strongly depends on the parameter values. For a slight modification of $U_g$, the final state alternates between $w \sim +1$ and $w \sim -1$ accounting for the BEC being located at the right or left trap, respectively. To understand the origin of this behavior, we plot in Fig.~4 the detailed dynamics for $U_g =1.7 \Omega_0$ corresponding to case (ii) in Fig.~2(b). At $\Omega_0 t=1374$, the solution that the system is adiabatically following is annihilated with an unstable one through a saddle-node bifurcation, i.e., case SNB42 previously described. At this point, the system splits into two non-degenerate components (being the largest one the energetically closest stable solution) and starts to oscillate at the frequency corresponding to the energy separation of the two solutions. At $\Omega_0 t=1588$, the system reaches a bifurcation point of the PB24 type. This bifurcation yields two energetically degenerated stable solutions corresponding to the two final states $w=\pm1$. Thus, at $\Omega_0 t=1588 $, the largest component of the system chooses one out of the two new stable solutions. We have numerically verified that the selected solution strongly depends on the previous oscillatory dynamics.   

\begin{figure}[t]
\includegraphics[scale=1.2]{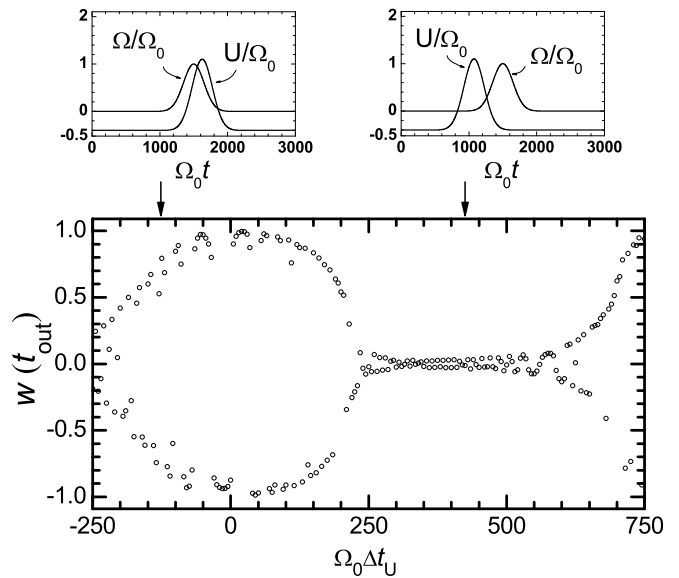}
\caption{\label{fig:scheme} Adiabatic splitting of a BEC. Population difference $w(t_{out})$ at the end of the BEC splitting process as a function of the temporal delay $\Delta t_U$.  The temporal variation of the non-linear interaction and the tunneling rate is shown in the two upper figures for $\Delta t_U=-125 \Omega_0^{-1}$ (left) and $\Delta t_U=425 \Omega_0^{-1}$ (right). Initial conditions: $w(t_{in})=-1$ and $u(t_{in})=v(t_{in})=0$. Parameter setting: $U_g=-0.4\Omega_0$, $U_0=1.5\Omega_0$, $ \sigma_U=\sigma_{\Omega}=212.8\Omega_0^{-1}$, $t_{\Omega}=1500\Omega_0^{-1} $ and $\epsilon_g=\epsilon_0=n_{\epsilon}=n_U=0$. The adiabatic splitting succeeds in the parameter region $\Omega_0\Delta t_{U} \simeq (250,600)$.
 }
\end{figure}
 
For $\left| U_g \right| >2.07 \Omega_0$ corresponding to regions $a$ and $e$ of Fig.~2(b), there are four stationary solutions during the whole dynamics that do not present any bifurcation. 
In this case, the non-linear coupling forces the adiabatic evolution from $\theta(t_{in})=0$ to $\theta(t_{out})=0$ instead of evolving to the desired $\theta(t_{out})=\pi /2$. Then, the initial $(u,v,w)=(0,0,-1)$ and the final state of the system coincide. 

Finally, for $-2.07\Omega_0<U_g<-0.12 \Omega_0$, region $b$ in Fig.~2, the energy eigenstate to be followed reaches, during the ramping down of the tunneling interaction $\Omega(t)$, a pitchfork bifurcation of the PB24 type. After this bifurcation point, the two stable stationary solutions are energetically degenerated and, therefore, the system splits into a non-oscillatory combination of them that for $U_g=-0.12 \Omega_0$ ($U_g=-2.07\Omega_0$) yields $w (t_{out}) = 0$ ($w (t_{out}) = -1$).

Up to now, we have discussed the possibility to adiabatically split a BEC by fixing $U$ and time varying $\epsilon (t)$ and $\Omega (t)$. Note, however, that  it is also possible to adiabatically split the BEC by appropriately time varying the non-linear interaction $U(t)$ and the tunneling rate $\Omega(t)$ with a fixed energy bias. Thus, for $U_g=-0.4\Omega_0$, $U_0=1.5\Omega_0$, $\epsilon = 0$ and the rest of the parameters given in the figure caption, Fig.~5 shows $w(t_{out})$ as a function of the time delay between the modulation of the tunneling rate and the non-linear interaction. We again obtain a plateau of parameter values, $\Omega_0\Delta t_{U} \simeq (250,600)$, where the adiabatic splitting, corresponding to $w (t_{out}) \sim 0$, becomes successful. A detailed analysis of the temporal dynamics reveals that, as in Fig.~2, the plateau corresponds to the case where the selected solution reaches during its dynamics a single bifurcation point of the PB42 type. 

\begin{figure}[t]
\includegraphics[scale=1.2]{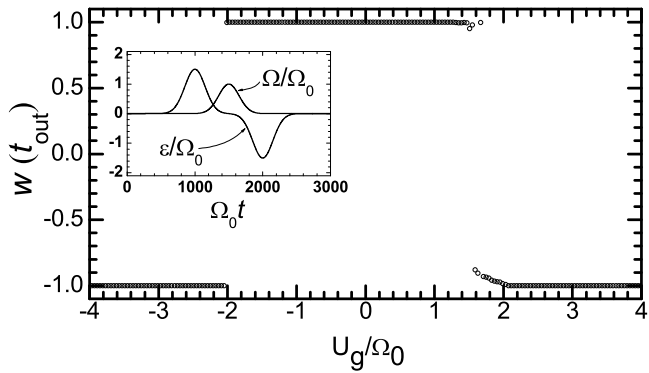}
\caption{\label{fig:scheme} Adiabatic transport of a BEC. Population difference $w(t_{out})$ at the end of the BEC transport process as a function of the non-linear interaction parameter $U_g$. The temporal variation of the energy bias and the tunneling rate is shown in the inset.
Initial conditions: $w(t_{in})=-1$ and $u(t_{in})=v(t_{in})=0$. Parameter setting: $\epsilon_0=1.5\Omega_0$, $ \sigma_{\epsilon}=\sigma_{\Omega}=212.8\Omega_0^{-1}$, $\Delta t_{\epsilon}=500\Omega_0^{-1}$, $t_{\Omega}=1500\Omega_0^{-1} $, $n_{\epsilon}=-1$ and $U_0=\epsilon_g=n_U=0$. The adiabatic transport succeeds in the parameter region $U_g/\Omega_0 = (-2,1.4)$. 
}
\end{figure}

\subsection{Adiabatic Transport}

To adiabatically transfer the BEC from the left to the right trap, the mixing angle must be slowly varied from $\theta (t_{in}) = 0$ up to $\theta (t_{out}) = \pi$. Taking $n_{\epsilon} = -1$ and the rest of parameters as in Fig.~2, Fig.~6 shows the population difference $w (t_{out})$ at the end of the antisymmetric double STIRAP process (shown in the inset) as a function of the non-linear interaction $U_g$. Clearly, there is a wide parameter setting, the plateau $U_g/\Omega_0 = (-2,1.4)$, where the BEC transfer process takes places with a high fidelity. In the parameter region $U_g/\Omega_0 = (-2,0)$, the system starts at $w(t_{in})=-1$ following an energy eigenstate of the system that does not involve any bifurcation point. For $U_g/\Omega_0 = (0,1.4)$ the system crosses first a PB42 bifurcation to later on reaching a PB24 bifurcation point that, for these parameters, yields eventually $w(t_{out})=1$. In regions $U_g/\Omega_0 = (-4,-2)$ and $U_g/\Omega_0 = (1.4,4)$ the non-linearity yields $\theta (t_{out}) = 0$ forcing the initial and final state to be the same. 

\begin{figure}[t]
\includegraphics[scale=1.2]{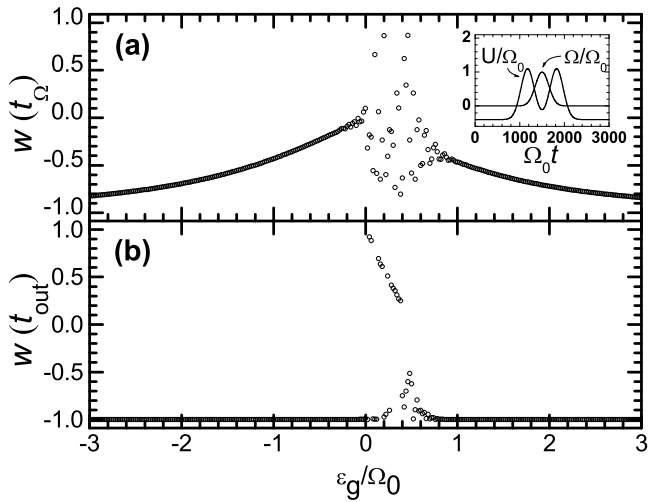}
\caption{\label{fig:scheme} Adiabatic self-trapping of a BEC. Population difference at (a) the intermediate time $t=t_{\Omega}$ and (b) at the end of the BEC transport process, as a function of the energy bias $\epsilon_g$.  The temporal variation of the non-linear interaction and the tunneling rate is shown in the inset for $U_0=1.5\Omega_0$. Initial conditions: $w(t_{in})=-1$ and $u(t_{in})=v(t_{in})=0$. Parameter setting: $U_g=-0.4\Omega_0$, $ \sigma_U=\sigma_{\Omega}=212.8\Omega_0^{-1}$, $\Delta t_U=325\Omega_0^{-1}$, $t_{\Omega}=1500\Omega_0^{-1} $, $n_U=1$ and $\epsilon_0=U_0=n_{\epsilon}=0$. The adiabatic self-trapping state does not succeed in the parameter region $\epsilon_g/\Omega_0 = (0,0.8)$.}
\end{figure}

\subsection{Adiabatic Self Trapping}

It is also possible to completely inhibit the BEC transport between the two wells by means of the matter wave analogue to the double STIRAP process (see inset of Fig.~7). Starting with the BEC located at the left trap, $w (t_{in})=-1$, we show in Fig.~7 the population difference at the time when the tunneling rate is maximum, $w (t_{\Omega})$, and at the end of the double-STIRAP sequence, $w (t_{out})$, as a function of the energy bias $\epsilon_g$. The adiabatic self-trapping process succeeds for a wide domain of parameter values, except for the region $\epsilon_g/\Omega_0 = (0,0.8)$. In this last region, the 
solution to be followed reaches during its dynamics a saddle-point bifurcation while, for the rest of values of the energy bias, the selected solution does not involve any bifurcation point.   

\section{Conclusions}

By means of a dark variable that results from the combination of the population imbalance and the spatial atomic coherence, we have investigated in detail the adiabatic dynamics of a BEC in a double well potential in the framework of the two-level approximation. We have shown that it is possible to robustly split, transport or trap a BEC by  appropriate temporal variation of either the energy bias or the non-linear interaction together with the tunneling rate. All these proposals have been studied from a non-linear dynamics perspective deriving the stationary solutions of the system, evaluating their linear stability, and discussing the bifurcation scenarios. For the eventual implementation of the techniques here discussed, we want to highlight the recent work by M. Bauer \textit{et al}. \cite{Bauer09} reporting an accurate control on magnetic Feshbach resonances by means of laser light.  

In order to check the validity of the two-mode approximation, we have numerically integrated the one-dimensional Gross-Pitaevskii (GP) equation for the non-linear adiabatic splitting of a BEC in a double-well potential, obtaining the characteristic 'plateau' of the STIRAP protocol shown in Fig. 2(b). However, a detailed numerical investigation of the GP equation still remains to be performed to validate that the matter wave non-linear STIRAP techniques here derived in the two-mode approximation could be used for matter wave interferometry or coherent transport of a BEC. In performing this analysis, the global landscape provided by the results of the two-mode approximation should be a useful roadmap.

Although it is out of the scope of the present paper, it would be very interesting also to extend the present non-linear matter wave STIRAP techniques to the second-quantization formalism \cite{secondq}. Within this formalism, we could elucidate whether the adiabatic splitting of the BEC results in a macroscopic superposition where the entire condensate localizes in one of the wells, i.e. in a NOON state \cite{NOON}, or in a perfect 50\% spatial splitting of all the atoms. Note that in both cases the final population difference reads $w (t_{out})$=0. 

\bigskip
\begin{acknowledgments}

We acknowledge support from the Spanish Ministry of Education and Science 
under contracts FIS2008-02425 and CSD2006-00019, and for the "Juan de la Cierva" postdoctoral fellowship (C.~O.). 
Support from the Catalan Government under contract SGR2009-00347 is also acknowledged. 
We thank A.~Benseny, G.~Birkl, Y.~Loika, and G.~Orriols for valuable and clarifying discussions. 

\end{acknowledgments}

\appendix
\section{LINEAR STABILITY ANALYSIS}

The stationary solutions of the density matrix equations~(7) take the form $\left\{ u^{ss},v^{ss}=0,w^{ss} \right\} $ with
$u^{ss}/w^{ss}=2\Omega / (\epsilon + U w^{ss})$ and $(u^{ss})^2+(w^{ss})^2=1$.
Let us consider now a perturbation around any of these solutions in the form of:
\begin{eqnarray}
u(t) &=& u^{ss}+ \delta u e^{\lambda t}  \\
v(t) &=&  \delta v e^{\lambda t} \\
w(t) &=& w^{ss}+ \delta w e^{\lambda t} 
\end{eqnarray}
where the real part of $\lambda$ determines the stability of the solution. 
The stability is governed by the equation $(\hat{M}- \lambda \hat{I})\delta  \vec{s} = 0$ 
where $\delta \vec{s} = ( \delta u, \delta v, \delta w )^{T}$ and $\hat{M}$ is the linear stability matrix defined as:
\begin{equation}
\hat{M} \equiv \left( 
\begin{array}{ccc}
0 & -\left( \epsilon + U w^{ss} \right) & 0 \\ 
\left( \epsilon + U w^{ss} \right) & 0 & U u^{ss}-2\Omega \\ 
0 & 2\Omega & 0
\end{array}
\right)
\end{equation}
Solving the secular equation  $ {\rm det}({\hat{M}- \lambda \hat{I}})=0 $, one obtains the following three eigenvalues:
\begin{eqnarray}
\lambda_{0} &=&0 \\
\lambda_{\pm} &=& \pm \sqrt{2\Omega (U u^{ss}-2\Omega)-(\epsilon+Uw^{ss})^2}
\end{eqnarray}
As expected from a conservative system, $\Sigma_{i=1}^3 \lambda_i = 0 $ for every stationary solution. One eigenvalue is equal to zero while the other two are either pure imaginary corresponding to a stable fixed point (or center), or pure real accounting for an unstable one (or saddle point).

\end{document}